\documentclass{amsart}

\newtheorem{theorem}{Theorem}[section]
\newtheorem{lemma}[theorem]{Lemma}

\theoremstyle{definition}
\newtheorem{definition}[theorem]{Definition}

\theoremstyle{remark}

\newcommand{\cA}{{\mathcal A}}
\newcommand{\cB}{{\mathcal B}}

\newcommand{\cH}{{\mathcal H}}

\newcommand{\cS}{{\mathcal S}}

\newcommand{\bC}{{\mathbb{C}}}
\newsymbol\boxtimes 1202

\numberwithin{equation}{section}

\newcommand{\vr}{\varrho}
\newcommand{\svr}{\vr^{1/2}}

\newcommand{\al}{\alpha}
\newcommand{\tom}{\widetilde{\Omega}}
\newcommand{\prep}{\pi_{\omega}^{\prime}}
\newcommand{\tprep}{\widetilde{\pi_{\omega}}}
\newcommand{\teha}{\widetilde{\cH}}

\begin{document}
\begin{center}
\vspace*{15mm}
{\LARGE ON QUANTUM CORRELATIONS AND POSITIVE MAPS
}\\
\vspace{2cm}
\textsc{{\large W{\l}adys{\l}aw A. Majewski}\\
Institute of Theoretical Physics and Astrophysics\\
Gda{\'n}sk University\\
Wita Stwosza~57\\
80-952 Gda{\'n}sk, Poland}\\
\textit{E-mail address:} \texttt{fizwam@univ.gda.pl}\\
\end{center}
\vspace*{4cm}
\noindent
\textsc{Abstract.}

We present a discussion on local quantum correlations and their 
relations with entanglement. We prove that vanishing coefficient of quantum correlations 
implies separability. The new results on locally decomposable maps which we
obtain in the course of proof also seem to be of independent interest.

\vspace{1cm} {\bf Mathematical Subject Classification}: Primary: 46L53, 46L60:
Secondary: 46L45, 46L30

\vspace{1cm}

\newpage
\section{Introduction}
Let $\cA_1$ and $\cA_2$ be $C^*$-algebras. For simplicity, we assume that either $\cA_1$
or $\cA_2$ is a nuclear $C^*$-algebra. This assumption is not particularly restrictive
as most $C^*$-algebras associated with physical systems have this property.
Moreover, the assumption leads to a unique construction of the $C^*$-tensor product
of $\cA_1$ and $\cA_2$. Let $\cA = \cA_1 \otimes \cA_2$. We write
$\cS(\cA)$ ($\cS(\cA_1)$, $\cS(\cA_2)$) for 
the set of all states on $\cA_1 \otimes \cA_2 \equiv \cA$ ($\cA_1$, $\cA_2$).
We define, for a state $\omega$ in $\cS(\cA)$  the restriction maps:
$$(r_1 \omega)(A) \equiv \omega(A \otimes {\bf 1}),$$
where $A \in \cA_1$ and
$$(r_2 \omega)(B) \equiv \omega({\bf 1} \otimes B),$$
where $B \in \cA_2$. 
Obviously,
 $r_i \omega$ is a state in $\cS(\cA_i)$, where $i=1,2$.
Next, take a measure $\mu$ on $\cS(\cA)$. Using the restriction maps one can define
measures $\mu_i$ on $\cS(\cA_i)$ in the following way: for a Borel subset
$F_i \subset \cS(\cA_i)$ we put
\begin{equation}
\mu_i(F_i) = \mu(r_i^{-1}(F_i)),
\end{equation}
where $i=1,2$.
Having measures $\mu_1$ and $\mu_2$, both originating
from the given measure $\mu$ on $\cS(\cA)$ one can define new measure $\boxtimes \mu$
on $\cS(\cA_1) \times \cS(\cA_2)$ which encodes classical correlations
between two subsystems described by $\cA_1$ and $\cA_2$ respectively (see \cite{M}). 
We first define $\boxtimes \mu$ for discrete measures
$\mu^d = \sum_i \lambda^d_i \delta_{\rho^d_i}$
with $\lambda^d_i \ge 0$, $\sum_i \lambda^d_i =1$, $\rho^d_i \in \cS(\cA)$.
$\delta_{\sigma}$ stands for Dirac measure. 
We introduce
$\mu^d_1 = \sum_i \lambda^d_i \delta_{r_1\rho^d_i}$
and $\mu^d_{2} = \sum_i \lambda^d_i \delta_{r_2\rho^d_i}$. Define
\begin{equation}
\label{gwiazdka2}
\boxtimes \mu^d = \sum_i \lambda^d_i \delta_{r_1 \rho^d_i} \times 
\delta_{r_2 \rho^d_i}.
\end{equation}
Next, let us take an arbitrary measure $\mu$ in $M_{\phi}(\cS)$. 
Here, $M_{\phi}(\cS) = \{ \mu: \phi = \int_{\cS}\nu d\mu(\nu)\}$; i.e. the set of all 
Radon probability measures on $\cS(\cA)$ with the fixed barycenter $\phi$.
For the measure $\mu$, there exists
net of discrete measures $\mu_k$ such that $\mu_k \to \mu$ ($^*$-weakly).
Defining $\mu^k_1$ ($\mu_{2}^k$) analogously as $\mu_1$ ($\mu_{2}$
respectively), one has $\mu^k_1 \to \mu_1$ and $\mu^k_{2} \to \mu_{2}$ where the convergence
is taken in $^*$-weak topology. Then define, for each $k$,
$\boxtimes \mu^k$ as in (\ref{gwiazdka2}). 
One can verify that $\{ \boxtimes \mu^k \}_k$
is convergent to a measure on $\cS(\cA_1) \times \cS(\cA_2)$,
so taking the weak limit we arrive
to the measure $\boxtimes \mu$ on $\cS(\cA_1) \times \cS(\cA_2)$.
It follows easily that $\boxtimes \mu$ does not depend on the chosen approximation procedure.

The measure $\boxtimes \mu$ leads to the concept of degree of local (quantum) correlations
for $\phi \in \cS(\cA), a_1 \in \cA_1, a_2 \in \cA_2$, which is defined as
\begin{eqnarray}
d(\phi, a_1, a_2)& = & \inf_{\mu \in M_{\phi}(\cS(\cA))}
|\phi(a_1 \otimes a_2) \nonumber \\
&& - (\int \xi d(\boxtimes \mu)(\xi))(a_1 \otimes a_2)|. \nonumber
\end{eqnarray}

Recently,  
we have studied relations between the coefficient of quantum correlations 
and entanglement (cf \cite{M}). 
R. Werner has kindly pointed out that the proof of the statement  
saying that {\it $d(\phi; a,b,)=0$ for all $a \in \cA_1$, $b \in \cA_2$ and
a state $\phi$ on $\cA$ implies
separability of $\phi$} contains a gap (see Proposition 5.3 in \cite{M}). 
The aim of this letter is to give the proof of 
the properly amended statement (Theorem 4.3, Section 4). To this end we also give a 
generalization of St{\o}rmer theory of locally decomposable
maps (see Section 3) which seems to be of independent interest. 
All definitions and notations used here are taken from \cite{M}.

\section{Local separability 1.}
Assume $d(\phi; a,b) = 0$ for all $a \in \cA_1$, $b \in \cA_2$
and for a state $\phi$ on $\cA$. Then 
as $\mu \mapsto (\int \xi d(\boxtimes \mu)(\xi))(a_1 \otimes a_2)$
is $^*$-weak continuous,
there exists a measure
$\mu \in M_{\phi}(\cS)$ (Radon probability measures on $\cS(\cA_1 \otimes \cA_2)$ 
with barycenter $\phi$) such that
\begin{equation}
\label{a}
\phi(a \otimes b) = \int_{\cS(\cA_1) \times \cS(\cA_2)} \xi d(\boxtimes \mu) (a \otimes b).
\end{equation}
Using the Riemann approximation property of the classical measure one has
\begin{equation}
\label{b}
\phi(a \otimes b) = \lim \sum_i \lambda_i(a,b)\xi^{(1)}_i(a) \xi^{(2)}_i(b),
\end{equation}
where $\lambda_i(a,b)$ are non-negative numbers, depending on $a$ and $b$, $\sum_i 
\lambda_i(a,b) = 1$ and states $\xi_i^{(1)}$ ($\xi_i^{(2)}$) are defined on
$\cA_1$ (on $\cA_2$ respectively) and depend on the chosen element $a \otimes b$.

\begin{definition}
Let a state $\phi$ on $\cA_1 \otimes \cA_2$ have a representation
of the form (\ref{a}) with the measure $\mu$ depending on the chosen element $a \otimes b$.
Such state will be called locally separable.
\end{definition}

In other words, one can say that if the coefficient of quantum correlations for a state $\phi$ 
vanishes on $a \otimes b$ then the state $\phi$ is locally separable. 
Now we wish to examine the property of local separability. Let us begin with 
a particular case: assume that $a$ is a normal element of $\cA_1$ while $b$ is arbitrary one
in $\cA_2$. Let $\phi \in \cS(\cA_1 \otimes \cA_2)$.  We observe that
\begin{equation}
\phi(a \otimes b) = \phi|_{\cA_1^0 \otimes \cA_2^0} (a \otimes b),
\end{equation}
where $\phi|_{\cA_1^0 \otimes \cA_2^0}$ is the restriction of $\phi$ 
to the subalgebra $\cA_1^0 \otimes \cA_2^0 \subset \cA_1 \otimes \cA_2$.
Here, $\cA_1^0 $ is the abelian $C^*$-algebra 
generated by $a$ and $\bf 1$ ($a$ was normal!) while $\cA_2^0$ is 
the algebra, in general non-commutative, generated by $b$ and $\bf 1$.
But in such case, each state in $\cS(\cA_1^0 \otimes \cA_2^0)$ is a separable one. 
Moreover, $\phi$ has the decomposition depending on $a$ and $b$. 
However, we 
wish to stress: the assumption of normality for $a$ was crucial. Namely, taking
an arbitrary $a$ and $b$,  the condition of vanishing of coefficient $d$ implies
the uniformity of decomposition with respect to hermitian and antihermitian
part of $a$ in $a \otimes b$.
In that context it is worth adding that by the genuine separability we understand 
decomposition of type (\ref{a}) which is uniform with respect to elements of algebra $\cA$.

\smallskip

To show that $d(\phi, \cdot) = 0$ can imply separability, 
we will use another property of entangled states. Namely, 
one of the intriguing features of non-separable states is their complicated behaviour
under transformations by positive maps. To be more precise, one is interested
in inspection of the functional  $\phi \circ \alpha \otimes id_2 (\cdot)$,
where $\phi$ is a state on $\cA = \cA_1 \otimes \cA_2$, $\alpha: \cA_1 \to \cA_1$
is a linear, unital positive map while $id_2$ is the identity map on $\cA_2$.
To proceed with answering this question we need a description of locally decomposable maps and a 
modification of definition of coefficient of quantum correlations which will be given
in the next sections.

\section{Locally decomposable maps}
This section is a fairly straightforward generalization of the St{\o}rmer 
concept of local decomposibility; see Definition 7.1 as well as Lemma 7.2 and Theorem
7.4 in  \cite{S}.
\begin{definition}
Let $\alpha$ be a linear positive map of a $C^*$-algebra $\cA$ into 
$\cB(\cH)$, $\cH$ being a Hilbert space. The map $\alpha$ is locally decomposable if for 
each normal state $\phi(\cdot) \equiv Tr\varrho(\cdot)$ on $\cB(\cH)$
there exists a Hilbert space $\cH_{\varrho}$, and a linear map $V_{\varrho}$
of $\cH_{\varrho}$ into $\cH_0 = <\cB(\cH) \svr>^{cl}$ with property $||V_{\vr}|| \le M$
for all $\vr$ and a $C^*$-homomorphism $\pi_{\vr}$ of $\cA$ into 
$\cB(\cH_{\vr})$ such that
$$V_{\vr} \pi_{\vr}(a) V^*_{\vr} \svr = \alpha(a) \svr, $$
for all $a \in \cA$.
\end{definition}

We will need

\begin{lemma}
Let $\cA$ be a $C^*$-algebra, $\cH$ a Hilbert space, and $\al$ a positive unital 
linear map of $\cA$ into $\cB(\cH)$. If $\vr$ is a density matrix on $\cH$ defining
a normal state $\phi$ on $\cB(\cH)$ then there is a $^*$-representation $\pi$ of $\cA$ as 
$C^*$-algebra on a Hilbert space $\cH_{\pi}$, a vector $\Omega_{\pi} \in \cH_{\pi}$
cyclic under $\pi(\cA)$, and a bounded linear map $V$ of the set 
$\{ \pi(a) \Omega{_\pi}; a \in \cA, a = a^*\}^{cl}$  into $\cH_{\vr} = <\al(a)\svr; a \in \cA>^{cl}$ such that
$$V\pi(a)V^* \svr = \al(a) \svr,$$
for each self-adjoint $a$ in $\cA$.
\end{lemma}
\begin{proof}
Let $\omega(\cdot) = Tr \vr \al(\cdot)$.
Denote by $\pi_{\omega}$ the $^*$-representation of $\cA$ induced by $\omega$ on $\cH_{\omega}$
and let $\Omega$ be a cyclic vector for $\pi_{\omega}(\cA)$ in $\cH_{\omega}$
such that $\omega(\cdot) = (\Omega, \pi_{\omega}(\cdot) \Omega)$.
For selfadjoint $a \in \cA$, define $V\pi_{\omega}(a) \Omega = \al(a) \svr.$
The set $\{\pi_{\omega}(a) \Omega; a = a^*, a \in \cA \}^{cl}$
is a real linear subspace of $\cH_{\omega}$ whose complexification is dense
in $\cH_{\omega}$. If $\pi_{\omega}(a) \Omega =0$ then
$$0 = (\pi_{\omega}(a^2) \Omega, \Omega) = \omega(a^2) = Tr \vr \al(a^2)
\ge Tr \vr (\al(a))^2 \ge 0.$$
Hence $\al(a) \svr =0$. It follows that $V$ is well defined and linear. Note that
$$V \pi_{\omega}({\bf 1})\Omega = V \Omega = \al({\bf 1}) \svr, $$
and that
$$(V^* \svr, \pi_{\omega}(a) \Omega) = (\svr, V \pi_{\omega}(a) \Omega)
= (\svr, \al(a) \svr) = \omega(a) = (\Omega, \pi_{\omega}(a) \Omega),$$
for any self-adjoint $a \in \cA$. Thus $V^* \svr = \Omega$ and
$V \pi_{\omega}(a) V^* \svr = \al(a) \svr$ for each self-adjoint $a \in \cA$. Moreover
$$||\al(a) \svr ||^2 =  (\al(a)^2 \svr, \svr) \le (\al(a^2) \svr, \svr)
= \omega(a^2) = ||\pi_{\omega}(a) \Omega||^2,$$
so that $||V|| \le 1$ and with the identification, 
$\pi = \pi_{\omega}$, $\Omega = \Omega_{\pi}$, the proof is complete.
\end{proof}

Now, we recall (see Lemma 7.3 in \cite{S}): If $\al : \cA \to \cB(\cH)$ 
is unital, positive map then
\begin{equation}
\label{jeden}
\al(a^*a + a a^*) \ge \al(a^*) \al(a) + \al(a) \al(a^*),
\end{equation}
for all $a \in \cA$. 
Lemma 3.2 and the inequality (\ref{jeden}) lead to

\begin{theorem}
Every unital positive linear map of a $C^*$-algebra $\cA$ into $\cB(\cH)$
is locally decomposable.
\end{theorem}
\begin{proof}
Let $\vr$, $\omega$ and $\pi_{\omega}$ be as in Lemma 3.2.
Define $\prep$ in terms of the right kernel as a $^*$-anti-homomorphism (i.e.
$<a,b> = \omega(ab^*)$, ${\rm I}_{\omega} = \{a; <a,a>=0 \}$, $\prep(c) (a + {\rm I}_{\omega})
=ac + {\rm I}_{\omega}$) 
of $\cA$ on the Hilbert space $\cH_{\omega}^{\prime}$
and let $\tprep = \pi_{\omega} \oplus \prep$. Let $\teha$
be the Hilbert space $\cH_{\omega} \oplus \cH_{\omega}^{\prime}$ with
the inner product
$$(z \oplus z^{\prime}, y \oplus y^{\prime}) = 1/2[(z,y) +<z^{\prime}, y^{\prime}>],$$
where $y,z \in \cH_{\omega}$, $y^{\prime}, z^{\prime} \in \cH_{\omega}^{\prime}$.
$\tprep$ is a $C^*$-homomorphism of $\cA$ into $\cB(\teha)$. With $\Omega$ and $\Omega^{\prime}$
the vacuum vectors of $\omega$ for $\pi_{\omega}$ and $\pi_{\omega}^{\prime}$ respectively, let
$\tom = \Omega \oplus \Omega^{\prime}$.
Define a map $V^{\prime}$ of the linear submanifold $\tprep(\cA) \tom$ of $\teha$
into $<\al(\cA)\svr>^{cl}$ by
$$V^{\prime} \tprep(a) \tom = \al(a) \svr,$$
for each $a \in \cA$. Note that if $\tprep(a)\tom =0$
then $\pi_{\omega}(a) \Omega =0 = \prep(a) \Omega^{\prime}$.
Thus
$$\pi_{\omega}(a^*) \pi_{\omega}(a) \Omega = \pi_{\omega}(a^*a) \Omega = 0 = 
\prep(a^*) \prep(a) \Omega^{\prime} = \prep(aa^*) \Omega^{\prime},$$
so that $\omega(aa^*) = 0 = \omega(a^*a)$. Thus by \ref{jeden}
$$0 = ((\al(a^*a)+\al(aa^*)) \svr, \svr) \ge ((\al(a^*) \al(a) + 
\al(a) \al(a^*)) \svr, \svr) \ge 0.$$
Hence $\al(a) \svr = 0$. Consequently, $V^{\prime}$ is well defined and linear. Moreover,
$$||V^{\prime}|| = sup \{ ||\al(a) \svr|| : ||\tprep(a) \tom|| = 1 \}
= sup \{ ||\al(a) \svr || : ||\pi_{\omega}(a) \Omega \oplus 
\pi^{\prime}_{\omega}(a) \Omega^{\prime}||^2 =1 \}$$
$$= sup \{ ||\al(a) \svr || : ((\al(a^*a) + \al(aa^*))\svr, \svr) = 2 \}.$$
By \ref{jeden}, if $(\al(a^*a +aa^*) \svr, \svr) = 2$
then $((\al(a^*) \al(a) + \al(a) \al(a^*)) \svr, \svr) \le 2$. Hence $||\al(a) \svr||^2 \le 2$. 
Consequently $||V^{\prime}|| \le 2^{{1/2}}$.

We extend $V^{\prime}$ by continuity to all of the subspace $\teha_0 = <\tprep(\cA) \tom>^{cl}$
and call the extension $V^{\prime}$. Define the linear map of $\teha$ into $<\al(\cA) \svr>^{cl}$ 
in the following way:
$V$ restricted to $\teha_0$ equals $V^{\prime}$ and $V$ 
restricted to orthocomplement of $\teha_0$ is equal to $0$.
Then $||V|| \le 2^{1/2}$. Moreover, repeating the corresponding argument given in the proof
of Lemma 3.2 one can show $(V^{\prime})^* \svr = \tom$ and this completes the proof.
\end{proof}

\section{Local separability 2.}
Having the notion of locally decomposable maps one might be tempted to
study local PPT (positive partial transposition) property, 
now without any restriction with respect to dimension. One can also study relations
between local separability and locally decomposable maps.
To proceed with these questions one should evaluate functionals
and study the coefficient $d(\cdot)$
on an arbitrary positive element of $\cA$. To this end we propose

\begin{definition}
Let $\phi$ be a state on $\cA = \cA_1 \otimes \cA_2$ and $A$ be an element
in $\cA$. The general coefficient of quantum 
correlations $d_0(\cdot)$ for $\phi$ and $A$ is defined as
\begin{equation}
\label{nowa def}
d_0(\phi, A) = \inf_{\mu \in M_{\phi}(\cS)} | \int_{\cS} \xi d\mu(\xi) (A) 
-  \int_{\cS_1 \times \cS_2} \xi d(\boxtimes \mu)(\xi) (A)|.
\end{equation}

\end{definition}

\vskip 1,5cm

To clarify this definition we recall that, by definition, $\mu_1$ and $\mu_{2}$
are probability measures on $\cS(\cA_1)$ and $\cS(\cA_2)$, respectively 
(they are basic ingredients of the definition of $\boxtimes\mu$; see 
Introduction or \cite{M}).
Consequently, $\boxtimes \mu$ is a probability measure on $\cS(\cA_1)
\times \cS(\cA_2)$. However, as $\cS(\cA_1)
\times \cS(\cA_2) \subset \cS$ is a measurable subset of $\cS$ one can consider
$\boxtimes \mu$ as a probability measure on $\cS$ supported by $\cS(\cA_1)
\times \cS(\cA_2)$. To summarize, $\int_{\cS_1 \times \cS_2} \xi d(\boxtimes \mu)(\xi)$
is a well defined element of $\cS(\cA)$. 
Therefore $\int_{\cS_1 \times \cS_2} \xi d(\boxtimes \mu)(\xi)(A)
\equiv \sum_i \int_{\cS_1 \times \cS_2} \xi d(\boxtimes \mu)(\xi)(a_i \otimes b_i)$
is also well defined ($A=\sum_i a_i \otimes b_i$ is a general element of $\cA$).

Obviously, the just given definition of $d_0(\cdot)$ is equivalent 
to that given for $d(\cdot)$  (cf \cite{M}) if one restrict oneself to simple tensors!
Moreover, it is worth noting that, in measure terms, separability of $\phi$ is equivalent to
$\boxtimes \mu \in M_{\phi}(\cS)$ (cf \cite{A}).

\vskip 1cm

Let us consider a state $\phi$ on $\cA$ such that $d_0(\phi, A) =0$ for some fixed $A \in \cA
\equiv \cA_1 \otimes \cA_2$, where $\cA_1, \cA_2$ are finite dimensional
$C^*$-algebras. This is the most important case 
considered within Quantum Information Theory.
The general case needs more complicated arguments based on approximation procedures 
and it will be not considered here. We also 
assume that $A \ge 0$ and we suppose that the measure $\mu$ appearing in the
condition $d_0(\phi, A) =0$ is finitely supported. This involves
no loss of generality, as there exist (finite) optimal decompositions (cf \cite{M}).
Then, there are states $\{ \phi^1_{A;i}\} \subset \cS(\cA_1)$, and
$\{ \phi^2_{A;i}\} \subset \cS(\cA_2)$ and non-negative numbers $\lambda_i(A)$,
$\sum_i \lambda_i(A) = 1$ 
such that:
$$\phi(A) \equiv \phi(\sum_{kl} a^*_k a_l \otimes
b^*_k b_l)  = \sum_i \sum_{kl} \lambda_i(A) \phi^1_{A, i}( a^*_k a_l)
\phi^2_{A, i}( b^*_k b_l).$$
Now, we are in position to analyse $\phi \circ \alpha \otimes id_2$ for a state $\phi$ on $\cA$
having $d_0(\phi, A) =0$ for all $A \in \cA$. Here, $\alpha$ is an arbitrary 
linear unital positive map on $\cA_1$; $\alpha : \cA_1 \to \cA_1$. 
Moreover, we put $A \ge 0$ 
and again observe that
\begin{eqnarray}
\label{aa}
(\phi \circ \alpha \otimes id_2)(A) 
= \sum_i \sum_{kl} \lambda_i(A) \phi_{A,i}^1(\alpha(a^*_k a_l)) 
\phi_{A,i}^2(b^*_k b_l) \nonumber \\
=  \sum_i \sum_{kl} \lambda_i(A) \phi_{A,i}^1(V_{\phi, i, A}^* \pi_{\phi, i, A}(a^*_k a_l)
V_{\phi, i, A}) \phi_{A,i}^2(b^*_k b_l),
\end{eqnarray}
where $\pi_{\phi, i, A}(\cdot)$ is a $C^*$-morphism (cf Section 3).

\vskip 1cm

Our first remark on (\ref{aa})  is that any $C^*$-morphism is, in fact, a sum of $^*$-morphism
and $^*$-antimorphism (cf \cite{Tak} or \cite{BR}). 
The second observation says that $\{a^*_ka_l \}_{kl} $
and $\{b^*_k b_l \}_{kl}$ are positive semidefined matrices with $\cA_1$ 
($\cA_2$)-valued entries (cf \cite{Tak}). Taking states $\varphi^1$ and 
$\varphi^2$ on $\cA_1$ and $\cA_2$ respectively, one gets positive semidefined matrices
$\{ \varphi^1(a^*_ka_l) \}_{kl}$ and $\{ \varphi^2(b^*_kb_l) \}_{kl}$
with entries in $\bC$. The next remark is that the Hadamard product of 
positive semidefined matrices is a positive semidefined matrix (cf \cite{Ha}). Finally, 
we recall that the transposition 
of a positive semidefined matrix with complex valued entries is again
positive semidefined. Taking all that into account one gets:

\smallskip

\begin{lemma}
\label{lemat}
Assume that the antimorphism in the decomposition of $\pi_{\phi, i, A}$ is composed
of a $^*$-morphism with transposition. Then, for any positive $A \in \cA$
$(\phi \circ \alpha \otimes id_2)(A)$ is positive. Hence, provided that
the assumption of this Lemma is satisfied, a state $\phi$ with $d_0(\phi, A)=0$ 
for any $A \in \cA$ is the separable one.
\end{lemma}

\smallskip

We have used the fact that only separable states are invariant (globally) with
respect to ``partially positive maps'' (see \cite{W}, \cite{P}, \cite{H} and \cite{MM}). 
It is well known that any antimorphism can be represented
as the composition of morphism and transposition (transposition
is an antimorphism of order two, while the composition
of two antimorphisms leads to a morphism). 
Thus, the assumption of Lemma \ref{lemat} is always satisfied. As a conclusion one has
that the condition $d_0(\phi,A)=0$ for any $A\in \cA$ is the
sufficient condition for separability of $\phi$. 
Hence, we got
\begin{theorem}
Assume $\cA$ is the tensor product of finite dimensional $C^*$-algebras $\cA_1$ and $\cA_2$.
Then, a state $\phi$ is separable if and only if $d(\phi; A) =0$ for any $A \in \cA$.
\end{theorem}
\begin{proof}
We have just proved, Lemma (\ref{lemat}), that $d_0(\phi;A)=0$ 
for all $A \in \cA$ implies separability of $\phi$.
Conversely, the definition of separability implies that the coefficient
$d_0$ is equal to zero (cf \cite{M}). This completes the proof.
\end{proof}
We want to close this section with an obvious remark 
that having a state $\phi$ with $d_0(\phi, A)=0$ 
for any $A \in \cA$ , the positivity of partial transformation is the sufficient condition 
for separability.

\section{ Acknowledgments}
The author would like to thank the organisers of
Conference on Quantum Probability and Infinite Dimensional Analysis, 
HPRN-CT-2002-00279, Greifswald Germany, in particular Michael Sch\"urmann,
and the participants for a very nice and interesting conference. He thanks Reinhard
Werner, Louis Labuschagne and Marcin Marciniak for useful discussions on separability and
quantum correlations.
He would like also to acknowledge the support of the KBN grant PB/1490/PO3/2003/25

\end{document}